\documentclass{article}
\begin{document}

\newcommand{\ra}{\rightarrow}
\newcommand{\ko}{K^0}
\newcommand{\be}{\begin{equation}}
\newcommand{\ee}{\end{equation}}
\newcommand{\bea}{\begin{eqnarray}}
\newcommand{\eea}{\end{eqnarray}}
\newcommand{\mtitle}[1]{\begin{center}{#1}\end{center}}
\newcommand{\mauthor}[1]{\begin{center}{\small\rm #1}\end{center}}
\newcommand{\address}[1]{\begin{center}{\small\it #1}\end{center}}
\newcommand{\presentedby}[1]{\begin{center}{Presented by #1}\end{center}}

\hfill October 23, 1996 

\hfill IFUM-543/FT

\hfill NBI-HE-96-63  

\hfill hep-th/9611013

\vskip0.8truecm

\mtitle{SCATTERING AMPLITUDES AND THE CPT THEOREM IN STRING THEORY}

\vskip0.5truecm

\mauthor{A. PASQUINUCCI}
\address{Dipartimento di Fisica, Universit\`a di Milano,
and INFN, sezione di Milano\\via Celoria 16, I-20133 Milano, Italy}

\centerline{\small\rm and}

\mauthor{K. ROLAND}
\address{Niels Bohr Institute, University of Copenhagen,
Blegdamsvej 17 \\ DK-2100 Copenhagen, Denmark}

\vskip0.5truecm

\centerline{Seminar given at the Conference {\it Gauge Theories, Applied
Supersymmetry}}
\centerline{{\it and Quantum Gravity}, Imperial College, London, July 1996}

\presentedby{A. PASQUINUCCI}

\vskip0.5truecm


\begin{abstract}
We discuss the role of the CPT transformation in first
quantized string theories, both on the world-sheet and in the space-time.
We explicitly show that the space-time
CPT theorem holds for all first-quantized
(perturbative) string theories in a Minkowski background of 
even dimension $D>2$.
\end{abstract}

\section{Introduction}
The Scattering Matrix \index{Scattering Matrix}
is a central object in String Theory, even more
than in Field Theory since at the moment we lack a full lagrangian
formulation of String Theory and we mostly work in a first-quantized
formalism. Indeed, roughly speaking a first-quantized perturbative
String model is built by giving a 2d Conformal Field Theory
(CFT), a GSO projection of the space-time spectrum (or a modular invariant
one loop partition function) and the Polyakov formula for the scattering
amplitudes.

Many interesting results have already been obtained by the explicit
computation of string scattering amplitudes. Very well known are the
results obtained in Field Theory by comparing the string scattering
amplitudes in the so-called ``field-theory limit'' and the
same scattering amplitudes directly computed in Field Theory.
Even so, a lot of work is still to be done. Indeed, up
to now mainly tree level (genus zero) scattering amplitudes and
one-loop (genus one) space-time bosonic scattering amplitudes have
been computed. In ref.~\cite{ammedm} an example was given of a
one-loop scattering amplitude involving external space-time fermions.

Beside the technical problems which arise in the computation of
scattering amplitudes involving space-time fermions and in the
computation of multi-loop scattering amplitudes, we also need
to understand better the general properties of the string scattering
amplitudes. For example we know that the string scattering amplitudes
have different analytical properties than the field theory scattering
amplitudes.
On the other hand, we do not know exactly how unitarity,
causality,~\footnote{The concept of causality in string theory is not
quite well understood.
Here we refer to the generally accepted definition for which
a scattered wave should not reach the detector before the incident
wave strikes the target.} CPT and other similar common and well-known
properties in Field Theory, manifest themselves at the level of
theta functions, prime forms and the other mathematical objects which
explicitly form the integrand of a string scattering amplitude.
Of course, the knowledge of this can also help us in doing the
explicit computations.

Here we will report on some results that we have obtained concerning
the role of the CPT \index{CPT}
transformation, on the world-sheet and in the
space-time, in first-quantized (perturbative) string theories.
We will first briefly review the definition of the CPT transformation and
the statement of the CPT theorem in Field Theory. We will then
introduce an explicit formulation of the String Scattering amplitudes
and discuss some of its properties, including the role of the CPT
transformation on the 2d world-sheet. We will finally formulate
a space-time CPT transformation and show that for first-quantized
string models on a Minkowski background the CPT theorem holds true.

\section{CPT in Field Theory}
In Field Theory the general properties on which the CPT theorem is based
are:~\cite{CPT}

\noindent $\bullet$ Lorentz invariance;

\noindent $\bullet$ The energy is positive definite and there exists a
Poincar\'{e}-invariant vacuum, unique up to a phase factor;

\noindent $\bullet$ Local commutativity, i.e.\ field operators at space-like
separations either commute or anti-commute.

These general properties imply the validity of the spin-statistics relation
(i.e.\ fields of integer (half-integer) spin are quantized with
respect to Bose (Fermi) statistics) in $D>2$.
Then the spin-statistics \index{Spin Statistics}
relation, together with Lorentz invariance,
imply the CPT theorem. Indeed it is quite easy to check at the level
of Lagrangian field theory that the spin-statistics relation and
Lorentz invariance imply the CPT theorem.

We formulate the CPT transformation as the combination of
Pauli's Strong Reflection (SR) \index{Strong Reflection}
transformation and Hermitean Conjugation  \index{Hermitean Conjugation}
(HC): C~$+$~P~$+$~T~ = ~SR~$+$~HC. The
CPT transformation is anti-linear and is implemented
by an anti-unitary operator $\Theta$.

Strong Reflection transforms any single field
$\phi(\vec{x},t)$ into $\phi(-\vec{x},-t)$, times an appropriate phase
factor. It also inverts the order of a product of operators (for this reason
it is a symmetry of the operator algebra only). The basic
scalar, fermionic and vector fields transform as follows:
\bea
& & \phi(x) \ \ra \ \phi(-x) \\
& & \psi(x) \ \ra \ \varphi_{_{\rm SR}} \gamma^{D+1} \psi(-x) \qquad
{\rm with}
\qquad (\varphi_{_{\rm SR}})^2 = -(-1)^{D/2} \nonumber\\
& & \overline{\psi}(x) \ \ra \ - \varphi_{_{\rm SR}}^* \overline{\psi}(-x)
\gamma^{D+1} \nonumber\\
& & \phi^{\mu} (x) \ \ra \ - \phi^{\mu} (-x)\ . \nonumber
\eea
Here the phase $\varphi_{_{\rm SR}}$ is fixed (up to a sign)
by requiring consistency of the SR transformation
with Hermitean Conjugation for real fields.

The formulation of CPT as the combination of SR and HC turns out to be
convenient because, as we shall see, it can be adapted even to string
theory.

As well known, the C~$+$~P~$+$~T transformation acts on a state
as~\footnote{where $p$ is the momentum, $\eta$ the helicity or its
generalization in higher dimensions, and $\{\lambda\}$ a set of
charges or other labels.}
\be
\vert \rho \rangle \ = \ \vert p, \eta, \{ \lambda \} \rangle \
\ra \ \vert \rho^{\rm CPT} \rangle \ = \ \varphi_{_{\rm CPT}} (\eta,
\{ \lambda \} ) \ \vert p, -\eta, \{  -\lambda \} \rangle \ ,
\ee
which, of course, guarantees that in field theory for each particle
there exists the corresponding anti-particle (with opposite helicity).

For a scattering amplitude the statement of SR/CPT invariance, i.e.\
the CPT theorem, becomes
\bea
&&\langle \rho_{_{1}};
\ldots;\rho_{_{N_{out}}};in  \vert S\vert\rho_{_{N_{out}+1}};
\ldots;\rho_{_{N}};in \rangle\ =
\label{eq:ftcpt} \\
&&\qquad\qquad\qquad
= \langle \rho_{_{1}}^{\rm CPT}; \ldots; \rho_{_{N_{out}}}^{\rm CPT}
;out \vert S^{\dagger} \vert\rho_{_{N_{out}+1}}^{\rm CPT};
\ldots;\rho_{_{N}}^{\rm CPT}; out \rangle^*  \nonumber\\
&&\qquad\qquad\qquad
= \ \langle\, \rho_{_{N}}^{CPT};
\ldots; \rho_{_{N_{out}+1}}^{CPT};in
\vert S\vert\, \rho_{_{N_{out}}}^{CPT} ;\ldots;  \rho_{_{1}}^{CPT};in\rangle
\ . \nonumber
\eea

\section{Scattering Amplitudes in String Theory}
We formulate the string theory scattering amplitudes (the so-called
Polyakov formula) \index{Polyakov Formula}
in the operator formalism, in a form which mimics the
field theory Lehmann-Symanzik-Zimmermann (LSZ) formula.~\cite{normaliz}
\index{Lehmann-Symanzik-Zimmermann Formula} We consider
string theories on a Minkowski background of even dimension
$D>2$ with metric ${\rm diag} (-1,1,\ldots,1)$.
We define the $T$-matrix element by
\bea
&&{ \langle \rho_{_{1}}, \dots , \rho_{_{N_{\rm out}}} \vert S \vert
\rho_{_{N_{\rm out}+1}}, \dots , \rho_{_{N}}
\rangle_{\rm connected} \over \prod_{i=1}^{N}
\left( \langle \rho_i \vert \rho_i \rangle \right)^{1/2} }
= \\
&& \qquad
i (2\pi)^D \delta^D (p_1 + \dots p_{_{N_{\rm out}}} -
p_{_{N_{\rm out}+1}} - \dots - p_{_{N}})
\prod_{i=1}^{N} (2 p^0_i V)^{-1/2}\ \times\nonumber\\
&&\qquad T(\rho_{_{1}}; \dots ; \rho_{_{N_{\rm out}}} \vert
\rho_{_{N_{\rm out}+1}}; \dots ; \rho_{_{N}} )  \ , \nonumber
\eea
where
$ \vert \rho \rangle = \lim_{\zeta,\bar{\zeta} \rightarrow 0}
{\cal W}_{\vert \rho \rangle} (z=\zeta,\bar{z}=\bar{\zeta})
\ \vert 0 \rangle$ and ${\cal W}_{\vert\rho\rangle} =
c \bar{c} {\cal V}_{\vert\rho\rangle}$ is a primary conformal
field of dimension $\Delta=\overline{\Delta}=0$.
To write a completely explicit formula for the $T$-matrix element,
we may consider a heterotic string model in
the RNS formalism (but the formula  can be quite easily generalized
to any other first-quantized (perturbative) string model).
Then
\bea
&&T (\rho_{_{1}}; \dots ; \rho_{_{N_{\rm out}}} \vert
\rho_{_{N_{\rm out}+1}};\dots;\rho_{_{N}} ) \ = \label{masterformula} \\
&&\qquad\qquad =
\sum_{g=0}^{\infty} (-1)^{g-1} C_g \ \int \left(\prod_{I=1}^{3g-3+N}
{\rm d}^2 m^I\right) \ \prod_{\mu=1}^g \left( \sum_{\alpha_\mu,\beta_\mu}
C_{\beta_\mu}^{\alpha_\mu} \right) \ \times\nonumber\\
&&\qquad\qquad\quad
\langle \left\vert\prod_{I=1}^{3g-3+N} (\eta_I\vert b)\prod_{i=1}^N c(z_i)
\right\vert^2 \left(\prod_{A=1}^{N_{\rm PCO}} \Pi(w_A)\right)\ \times
\nonumber\\
&&\qquad\qquad\quad {\cal V}_{\langle \rho_1 \vert}^{(q_1)}
(z_1,\bar{z}_1) \dots {\cal V}_{\vert \rho_{_N} \rangle}^{(q_N)}
(z_{_N},\bar{z}_{_N})\rangle  \nonumber
\eea
where $q_i$ is the superghost charge of the $i^{\rm th}$ vertex
operator and $\Pi$ is the picture changing operator, the total number
of which is given by $N_{\rm PCO} = 2g-2 -\sum_{i=1}^N q_i$.
(For more details on the notation, see ref.~\cite{normaliz}.)

We should mention that to formulate the scattering amplitude in a
Min\-kow\-ski background one has to build the 2d spin-fields
describing the space-time fermions with the required signature.
This has been done in ref.~\cite{mink}.

The next step in a careful formulation of this scattering amplitude is
the normalization of the amplitude itself and of all vertex operators.
In a first-quantized formalism, the normalization does not come automatically
as for example from the Dyson formula in Field Theory, and one has to
use some general physical principle to fix it. One may fix the overall
normalization $C_g$ at any genus by considering the scattering of two
gravitons in the Regge regime (high center of mass energy
and low transfer momentum) and require the eikonal resummation of the
amplitude.  Analogously, we fixed the
normalization of all vertex operators describing incoming and outgoing
particles by requiring that the universal part of the absorption/emission
at genus zero of a very low energy graviton
assumes the form required by the principle of equivalence.
The results are explictly given in ref.~\cite{normaliz}.

One key point in fixing the normalization of the vertex operators is
the relation between the vertex operator ${\cal W}_{\vert \rho
\rangle}$ describing an incoming
particle and the vertex operator ${\cal W}_{\langle \rho \vert}$
describing the same particle but outgoing.
In field theory this relation is quite simply given by Hermitean Conjugation,
but in string theory it turns out to be given by the CPT transformation
on the world-sheet:~\cite{wscpt}
\index{CPT on the World-Sheet}
\be
{\cal W}_{\langle\rho\vert}(z=\zeta,\bar{z}=\bar\zeta)\ = \
(-1)^{q+1} \left({\cal W}_{\vert\rho\rangle}
(z=\zeta^*,\bar{z}=\bar\zeta^*)\right)^{\rm WS-CPT}
\label{eq:cptinout}
\ee
where $q$ is the superghosts charge (integer for space-time bosons and
half-integer for space-time fermions).

The World-Sheet CPT transformation is defined on the sphere
as the combination WS-CPT~=~BPZ~$+$~HC, where BPZ is the
Belavin-Polyakov-Zamolodchikov (BPZ) transformation~\cite{BPZ}
\index{Belavin-Polyakov-Zamolodchikov Transformation}
$z\ra 1/z$. Since $z$ is related to cylindrical coordinates by
$z=\exp(\tau+i\sigma)$, BPZ is seen to map $(\tau,\sigma)$ into
$(-\tau,-\sigma)$. From this point of view, BPZ is very much like
SR. However, unlike SR, BPZ does {\em not\/} invert the order of
operators and for this reason WS-CPT differs from the ordinary CPT
transformation defined as CPT = SR $+$ HC. Even so, WS-CPT leads to a
World-Sheet CPT Theorem,~\cite{wscpt}
\index{World-Sheet CPT Theorem} which
follows immediately from the fact that the BPZ transformation is a
global conformal diffeomorphism on the sphere and which is therefore
valid for any conformal field theory on the sphere, even
for ghosts and superghosts (which do not satisfy spin-statistics)
and for spin-fields (which are non-local).

The world-sheet CPT transformation and the world-sheet CPT theorem
can be extended to higher genus surfaces by means of
sewing,~\cite{Sonodatwo} \index{Sewing of Riemann Surfaces}
the detailed procedure of which is described in ref.~\cite{wscpt}.
Here it is sufficient to recall that the WS-CPT theorem for 2d CFTs on
a genus $g$ Riemann surface, when applied to eq.~\ref{masterformula},
leads to the following {\em formal\/} hermiticity property of
the string scattering amplitudes:
\be
\left(T(\rho_N;\ldots\vert\ldots;\rho_1)\right)^*\ =\
T(\rho_1;\ldots\vert\ldots;\rho_N) \ .
\ee
The proof of this equation is only {\em formal\/} because the integral
over the moduli in eq.~\ref{masterformula}
is not always convergent, and the regularization of the
divergencies gives rise to the imaginary part of the scattering amplitude
required by unitarity. Several regularization procedures (of varying
generality) have already appeared in the
literature,~\cite{regular}~\cite{Hoker}
even if an analytical prescription similar to the Feynman $i\epsilon$
prescription has not yet appeared.~\footnote{But see ref.~\cite{Hoker} for
a first step in this direction.}

\section{The Space-Time CPT Theorem in String Theory}
Recently there has been some interest in the question of possible
space-time CPT non-conservation
in string theory. Indeed, some mechanisms have been proposed that would
lead to CPT-breaking effects that might be detected in the next generation of
experiments.~\cite{Pott}~\cite{Ellis} Moreover, the interest in studying
CPT in String Theory is also in understanding better the role of space-time
CPT in presence of quantum gravity.

Not too much
is known and published on the space-time CPT properties of string theory.
Sonoda~\cite{Sonoda} discussed and proved the space-time CPT
theorem at the level of string perturbation theory for ten-dimensional
heterotic strings in a Minkowski background.
Kostelecky and Potting~\cite{Pott} proved the {\em dynamical\/}
CPT invariance of the open bosonic and super string field theories,
formulated in flat backgrounds --but they also suggested a method
whereby CPT might be broken {\em spontaneously}, based on the possibility
of a CPT non-invariant ground state.

Here we will briefly show that the space-time CPT theorem holds true for
any first-quantized string theory on a Minkowski background in even
dimensions $D>2$.~\cite{uscpt}
\index{Space-Time CPT Theorem in String Theory}

As in lagrangian Field Theory, we assume Lorentz invariance and
the spin-statistics relation. More explicitly, we consider first-quantized
string models in a Minkowski background, where
any two vertex operators describing physical states of half-integer
space-time spin anti-commute, whereas any other pair of physical
vertex operators commute. This ensures that the $T$-matrix elements
given by eq.~\ref{masterformula}
have standard spin-statistics properties under the exchange of any
pair of external states.

We define the space-time CPT transformation as a map on the world-sheet,
i.e.\ on the conformal fields.
In a generic first-quantized string theory on a Minkowski
background in $D>2$ dimensions, the space-time coordinates are
described by $D$ bosonic {\em free\/} world-sheet fields $X_\mu$
and, when there is world-sheet supersymmetry, by their world-sheet
superpartners $\psi_\mu$. We define the string Strong Reflection
as the transformation which maps
\be
X^\mu \ra -X^\mu \qquad\qquad \psi^\mu \ra -\psi^\mu\ ,
\ee
leaves all other world-sheet fields invariant and does {\em not\/}
invert the order of the operators. For consistency the spin-fields
describing space-time fermions must transform as
\be
S_A \ra \varphi_{_{\rm SR}}
\ ( \Gamma^{D+1} )_{A}^{\  B} S_B \ .
\ee
The world-sheet action, BRST current and GSO projection conditions
are invariant under this
transformation.

In field theory, the CPT transformation is defined as the combination
of SR and HC. In string theory we define the space-time CPT transformation
as the combination of the string SR transformation defined above and the map
(\ref{eq:cptinout})
between an incoming and outgoing vertex operator, i.e.\ the world-sheet
CPT transformation: Space-Time CPT~=~string
SR~$+$~(${\cal W}_{\vert\rho\rangle}\ra{\cal W}_{\langle\rho\vert}$)
\be
{\cal W}_{\vert \rho \rangle} \ \ra \
{\cal W}_{\vert \rho^{\rm CPT} \rangle} \equiv
\left( {\cal W}_{\langle \rho \vert} \right)^{\rm SR} =
(-1)^{q+1} \left(\left( {\cal W}_{\vert \rho \rangle}\right)^{\rm WS-CPT}
\right)^{\rm SR}\ .
\ee
Using the hypothesis of space-time Lorentz invariance, it is possible to
prove the following identity on any genus $g$ Riemann surface~\cite{uscpt}
\be
\langle ({\cal O}_1)^{\rm SR} (z_1,\bar{z}_1) \ldots ({\cal
O}_N)^{\rm SR} (z_N,\bar{z}_N)
\rangle = (-1)^{N_{\rm FP}}\ \langle {\cal O}_1 (z_1,\bar{z}_1) \ldots {\cal
O}_N (z_N,\bar{z}_N)
\rangle
\ee
where $2N_{\rm FP}$ is the number of space-time spinorial fields.
Then it holds
\bea
&&T( \rho_{_{N}}^{\rm CPT}; \ldots ; \rho_{_{N_{\rm
out}+1}}^{\rm CPT} \vert \rho_{_{N_{\rm out}}}^{\rm CPT}; \ldots ;
\rho_1^{\rm CPT} )\ =\
\sum \int \langle(\ldots) {\cal W}_{\langle \rho_{_{N}}^{\rm CPT} \vert }
\ldots {\cal W}_{\vert \rho_1^{\rm CPT} \rangle} \rangle
\nonumber\\
&&\qquad\qquad\qquad
=\  \sum \int \langle (\ldots)
\left({\cal W}_{\vert \rho_{_{N}} \rangle
}\right)^{\rm SR} \ldots \left({\cal W}_{\langle \rho_1 \vert}
\right)^{\rm SR} \rangle \nonumber\\
&&\qquad\qquad\qquad   =\ (-1)^{N_{\rm FP}}
\sum \int \langle(\ldots) {\cal W}_{\vert \rho_{_{N}} \rangle }
\ldots {\cal W}_{\langle \rho_1 \vert} \rangle \nonumber\\
&&\qquad\qquad\qquad
=\ T(\rho_1; \ldots ; \rho_{_{N_{\rm out}}} \vert \rho_{_{N_{\rm
out}+1}}; \ldots ; \rho_{_N} )\ , \label{eq:cptproof}
\eea
where in the last line we used the hypothesis of the standard
space-time spin-statistics relation.
Eq.~\ref{eq:cptproof} is the proof of the space-time CPT theorem
(see eq.~\ref{eq:ftcpt}), once
we show that the space-time CPT transformation we have defined acts in
the correct way on the space-time labels. In other words, the question
is: Given $\vert\rho\rangle = \vert k,\eta,\{\lambda\}\rangle$, do we have
$\vert\rho^{\rm CPT}\rangle \sim \vert k,-\eta,\{-\lambda\}\rangle$?

Consider for example a particle carrying a $U(1)$ charge $\lambda$.
The charge $\lambda$ is the eigenvalue of the zero mode of an hermitean
Ka\v{c}-Moody current $J_\Lambda$. Now, by definition $J_\Lambda$
does not transform under string SR, but under world-sheet CPT
$J_\Lambda \ra - J_\Lambda$, so that under space-time CPT $\lambda$
correctly changes sign. This argument can be generalized and extended
also to the helicity, thus finally proving that the space-time CPT
transformation we defined does indeed act in the correct way on the 
space-time labels. 
This concludes the proof of the space-time CPT theorem
for first-quantized string theories in a Minkowski background of
even dimension $D>2$.

\section*{Acknowledgments}
A.P.\ is supported by MURST and by the EU Science Programs no.\
SC1*-CT92-0879 and no.\ CHRX-CT-920035.
K.R.\ is supported by the Carlsberg Foundation
and by the EU Science Program no.\ SC1*-CT92-0879.

\newcommand{\Journal}[4]{{#1}{\bf #2}, #3 (#4)}
\newcommand{\NCA}{\em Nuovo Cimento}
\newcommand{\NIM}{\em Nucl. Instrum. Methods}
\newcommand{\NIMA}{{\em Nucl. Instrum. Methods} A}
\newcommand{\NPB}{{\em Nucl.\ Phys.\ }B}
\newcommand{\PLB}{{\em Phys.\ Lett.\ }B}
\newcommand{\PRL}{\em Phys.\ Rev.\ Lett.\ }
\newcommand{\PRD}{{\em Phys.\ Rev.\ }D}
\newcommand{\ZPC}{{\em Z. Phys.} C}
\newcommand{\PR}{\em Phys.\ Rev.\ }
\newcommand{\AP}{\em Ann.\ Phys.\ }

\end{document}